\documentclass[twocolumn,prl,aps,floatfix,superscriptaddress,longbibliography]{revtex4-1}

\usepackage{amsmath,amscd,times,graphicx,amssymb}

\usepackage[utf8]{inputenc}
\usepackage{amsfonts}

\usepackage{braket}
\usepackage{siunitx}
\usepackage{chemformula}

\begin{document}

\author{Tobias Serwatka}
\affiliation{Department of Chemistry, University of Waterloo, Ontario, N2L 3G1, Canada}
\author{Pierre-Nicholas Roy}
\affiliation{Department of Chemistry, University of Waterloo, Ontario, N2L 3G1, Canada}
\email{pnroy@uwaterloo.ca}

\title{Quantum Criticality and Universal Behavior in Molecular Dipolar Lattices of Endofullerenes}

\begin{abstract}
Fullerene cages allow the confinement of single molecules and the construction of molecular assemblies whose properties strongly deviate from those of free species. 
In this work, we employ the density-matrix renormalization group method to show that chains of fullerenes filled with polar molecules (LiF, HF, and H$_2$O) can form dipole-ordered quantum phases. In symmetry broken environments, these ordered phases are ferroelectric, a property that  makes them promising candidates for quantum devices. We demonstrate that for a given guest molecule, the occurrence of these quantum phases can be enforced or influenced by either changing the effective electric dipole moment or by isotopic substitution. In the ordered phase, all systems under consideration are characterized by a universal behavior that only depends on the ratio of the effective electric dipole divided by the rotational constant. A phase diagram is derived and further molecules are proposed as candidates for dipole-ordered endofullerene chains.  
\end{abstract}

\maketitle

The development of molecular surgery enables the insertion of single atoms and small molecules in fullerene cages~\cite{murata2008surgery,kurotobi2011single,bloodworth2022synthesis}. 
One can observe phenomena absent in the bulk since the guest molecule is shielded from direct interactions with species outside the cage. 
For instance,  H$_2$@C$_{60}$~\cite{kohama2009rotational} and H$_2$O@C$_{60}$~\cite{suzuki2019rotational} show a Schottky anomaly in their heat capacity curve, and for H$_2$O@C$_{60}$, one can observe spin conversion times of several hours between the \emph{ortho} and \emph{para} isomers~\cite{beduz2012quantum,mamone2014nuclear}.  
Because of shielding, systems like N@C$_{60}$ or P@C$_{60}$ are promising candidates for quantum computing~\cite{yang2010quantum,zhou2022implementation}. 
Most experiments focus on crystalline C$_{60}$ where  cages form a simple cubic lattice at low temperatures~\cite{david1991crystal}. But an appealing feature of endohedral systems is that in principle,  single cages can be arranged in a variety of possible assemblies such as monolayers~\cite{shi1994well,gulino2005engineered,hou2022synthesis},
or linear chains in carbon nanotubes~\cite{biskupek2020bond}. 
This rich variety of structural motifs allows the formation of new phases of matter for the respective guest molecules because short range interactions, such as hydrogen bonds which define the bulk phase in the case of water, are suppressed. 
For polar molecules like H$_2$O or HF, this creates structures in which the guest molecules mostly interact with each other via electrostatic forces. Hence it is possible to physically realize a dipolar lattice by using fullerene cages as a molecular environment rather than relying on optical lattices \cite{cioslowski1992endohedral,opticallattice2015}.

In this Letter, we study the formation of dipole-ordered quantum phases in
 chains of endofullerenes with 
 polar guest molecules using density matrix renormalization group (DMRG) calculations \cite{white1992density,fishman2022itensor,serwatka2022ground}. We observe the occurrence of a quantum phase transition between disordered and dipole-ordered quantum phases and discuss physical parameters that determine the appearance of these phases and the conditions for the actual realization of molecular quantum critical systems.

The system Hamiltonian for $N$ sites is
\begin{align}
\hat{H} = \sum_{i=1}^{N}\left(\hat{T}^{i}_{\mathrm{trans}} + \hat{T}^{i}_{\mathrm{rot}}+\hat{V}^{i}_{\mathrm{guest-cage}}\right)+\sum_{i=1}^{N-1}\hat{V}_{\mathrm{dd}}^{i,i+1}.
\label{eq:hamiltonian}
\end{align} 
Because we focus on \SI{0}{\kelvin} properties, and due to the adiabatic separation between molecular vibrations and other degrees of freedom, the molecules are assumed to be in their vibrational ground state averaged geometry. Equation~\eqref{eq:hamiltonian} contains different kinetic ($\hat{T}$) and potential ($\hat{V}$) energy contributions that compete with each other. Their balance is key for the quantum phase. The center of mass translational kinetic energy operator is
\begin{align}
\hat{T}^{i}_{\mathrm{trans}} = -\frac{1}{2M}\Delta_{\mathrm{COM},i}~,
\label{eq:Ttrans}
\end{align}
where $M$ is the total mass of molecule $i$ and $\Delta_{\mathrm{COM},i}$ its Cartesian Laplacian.
The rotational kinetic energy is
\begin{align}
\hat{T}_{\mathrm{rot}}^{i} =\begin{cases}
\hfil B_{0}\hat{J}_{i}^2 &,\text{linear rotor}\\
A_{e}\hat{J}^{2}_{a,i}+B_{e}\hat{J}^{2}_{b,i}+C_{e}\hat{J}^{2}_{c,i} &,\text{asymmetric top}~,
\end{cases} 
\label{eq:Trot}
\end{align}
with rotational constants $B_{0}$ or $A_{e}, B_{e},C_{e}$ and angular momentum operators $\hat{\mathbf{J}}_{i}=(\hat{J}_{a,i},\hat{J}_{b,i},\hat{J}_{c,i})^\intercal$. 
We employ system-specific potential models for $\hat{V}^{i}_{\mathrm{guest-cage}}$, the interaction potential between  C$_{60}$  and the guest (see Supporting Information). 
The dipole-dipole-interaction between neighboring cages is
\begin{align}
\hat{V}_{\mathrm{dd}}^{i,i+1} = \mu^2_{\mathrm{eff}}\left(\frac{\hat{\bf e}_{i}\cdot\hat{\bf e}_{i+1}-3(\hat{\bf r}_{i,i+1}\cdot\hat{\bf e}_{i})(\hat{\bf r}_{i,i+1}\cdot\hat{\bf e}_{i+1})}{\hat{R}_{i,i+1}^3}\right)
\label{eq:Vdd}
\end{align}
where $\hat{\bf e}_{i}$ denotes the unit vector operator along the electric dipole moment of the guest at cage $i$ and $\hat{\bf r}_{i,i+1}$ is the unit vector along the connection of the centres of mass of molecules $i$ and $i+1$ and $\hat{R}_{i,i+1}$ is their distance. 
In Eq.~\eqref{eq:Vdd} we use the effective electric dipole moment $\mu_{\mathrm{eff}} = \frac{\mu_{0}}{\sqrt{\varepsilon_{r}^{\mathrm{cage}}}}$.

As revealed in recent studies~\cite{serwatka2022efield,serwatka2023qpt} for the free water chain, it is the competition between the rotational kinetic energy and the dipole-dipole interactions that dictates the formation of a specific quantum phase. 
Rotations favor delocalization over all orientations leading to a disordered quantum phase. 
The significance of these terms is determined by the rotational constants. The larger they are, the larger is the rotational kinetic and thus the stronger the tendency to delocalize molecular orientations. 
Translations also favor positional delocalization but due to , it is less important here. 
In contrast, dipole-dipole interactions act against that delocalization by favoring molecular alignment along the axis of interaction (the chain axis). 
This angular localization leads to a dipole-ordered quantum phase. 
In general, the dipole-dipole interaction depends on three parameters. The first is $R_{0}$, the distance of the centres of mass of the single cages.
Its value affects the range of the distance $\hat{R}_{i,i+1}$ in Eq.~\eqref{eq:Vdd} and has a lower bound dictated by the cage diameter. 
Here, we keep this distance fixed at \SI{10}{\angstrom} which is a reasonable value considering the experimentally known van-der-Waals gaps of 3-\SI{4}{\angstrom} between fullerene cages in carbon nanotubes~\cite{biskupek2020bond}. 
The other two parameters are $\mu_{0}$, the electric dipole moment of the free guest molecule and $\varepsilon_{r}^{\mathrm{cage}}$, the relative permittivity of the cage. They can be combined as one parameter $\mu_{\mathrm{eff}}$ even though they originate from two different sources. $\mu_{0}$ is defined by the specific molecule and its rovibronic state. 
This study only considers rigid molecules in the vibronic ground state, which is reasonable for systems at zero temperature. The relative permittivity $\varepsilon_{r}^{\mathrm{cage}}$ is determined by the specific cage. For C$_{60}$, experiments and theoretical studies show a strong screening of the dipole moment of the guest by a factor of about one fourth~\cite{meier2015electrical}. 
Employing different cages allows a change for $\varepsilon_{r}^{\mathrm{cage}}$ and thereby $\mu_{\mathrm{eff}}$. Very recent theoretical studies suggest that by using Be$_{36}$O$_{36}$ cages, one observes hardly any screening and ionic-bond nanocages even show an anti-screening effect that enhances the effective electric dipole moment~\cite{silvestrelli2022screening}. 
Similar effects might also be possible by using heterofullerenes or functionalized carbon cages~\cite{lu2001structural,li2009theoretical,dheivamalar2015density}.

\begin{figure}[!htb]
\includegraphics[width=\columnwidth]{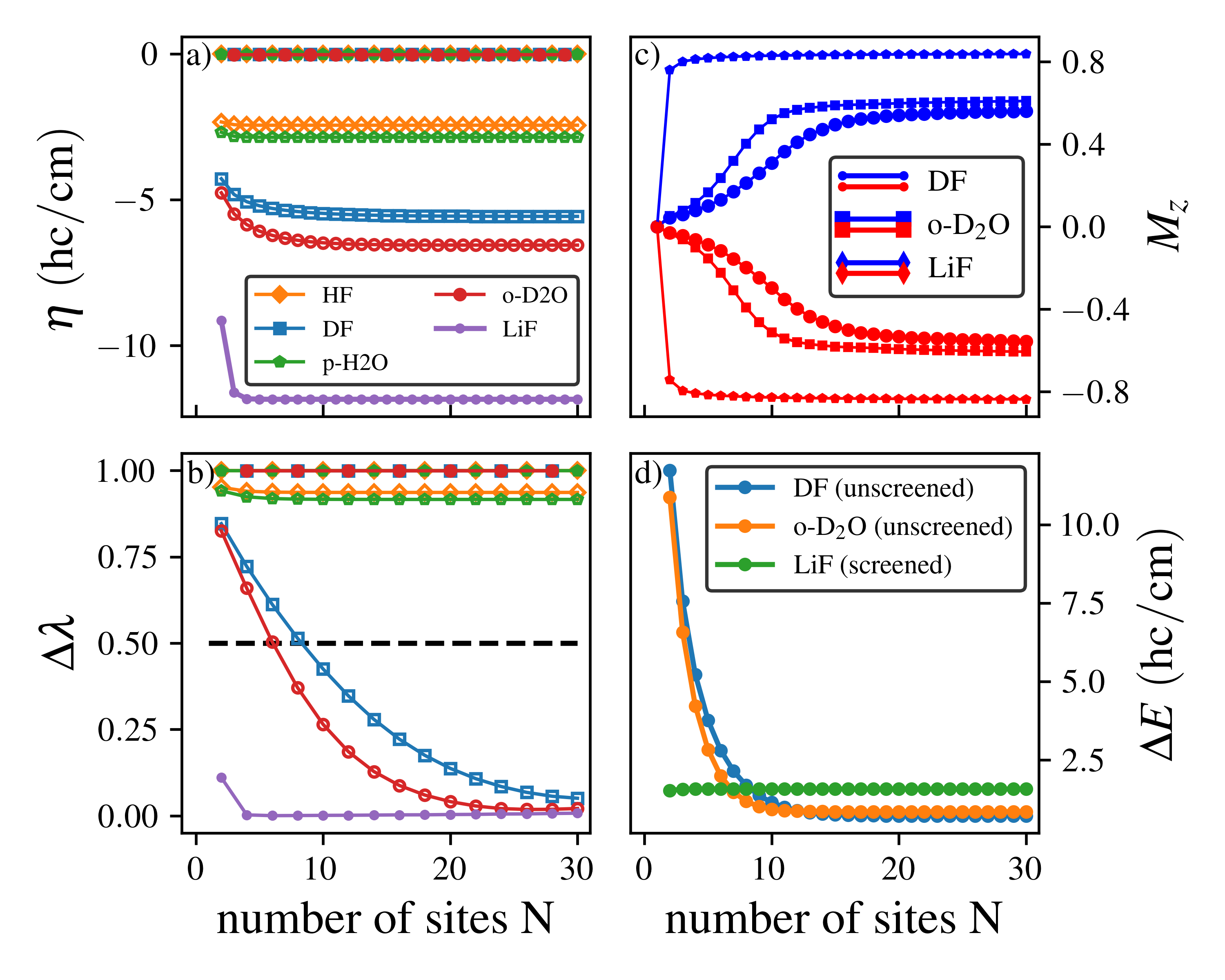}
\caption{a) Chemical potential and b) Schmidt gap for different guests. All curves in a) and b) are for molecules with screened (filled markers) and unscreened dipole moments (empty markers). For clarity, only every second point is shown for screened species except for LiF. c) Axial polarization for three different guests forming a dipole-ordered phase. The axial polarization is scaled by $N$ and by $\mu_0$ for  each guest. d) Fundamental energy gap for three guests forming a dipole-ordered phase. A small axial electric field $\vec{F}$ with $|\vec{\mu}\vec{F}|=\SI{0.5}{hc\per\centi\metre}$ is applied to the edge cages for c) and d) .}
\label{fig:sites}
\end{figure}

We study the impact of the mass distribution and the effective dipole moment on the formation of ordered quantum phases for three guest: LiF, HF and H$_2$O (see Supporting Information for molecular parameters). We only consider the p-H$_2$O (singlet spin) ground state.
The chemical potential $\eta$ is presented in Fig.~\ref{fig:sites}a) to highlight stability  and  convergence to the bulk limit.
In general, one can state that the higher $\mu_{\mathrm{eff}}$ and the smaller the rotational constants, the higher the gain in binding energy associated with  adding a new cage. 
LiF has the largest chemical potential in magnitude due to its strong dipole moment that favors dipolar bonding to neighboring cages. In contrast, for HF and p-H$_2$O there is hardly any bonding leading to an almost vanishing chemical potential. That is because in addition to a smaller $\mu_{\mathrm{eff}}$, the rotational constants are one order of magnitude larger than the rotational constant of LiF resulting in a weak angular localization. 
There are two possible strategies to increase that bonding for a given molecular species. The first is to increase the dipole-dipole interaction by increasing $\mu_{\mathrm{eff}}$. If one assumes a transparent cage that has no screening but a similar guest-host interaction one can perform calculations with the unscreened dipole moments of 1.83~D~\cite{lovas2005nist} and 1.86~D~\cite{shostak1991dipole} for HF and H$_2$O respectively. 
Fig.~\ref{fig:sites}a) shows that the larger dipole-dipole interactions leads to a smaller chemical potential and thus a stronger bonding. However, it is still much weaker than for LiF since the rotational constants of HF and H$_{2}$O are still an order of magnitude larger compared to LiF. To strengthen the bonding further, a second strategy can be employed. This approach focusses on actively decreasing the rotational terms by decreasing the rotational constants. That can be accomplished by isotopic substitution of hydrogen by deuterium in HF and H$_2$O. This substitution leads to much stronger bonding and a chemical potential that is in magnitude more than two times larger than for the hydrogen isotopologues (see Fig.~\ref{fig:sites}a)). In addition to stability, the chemical potential also shows how quickly the system reaches the bulk limit. Fig.~\ref{fig:sites}a) shows that all cases reach this behavior rather quickly. 
For LiF, HF and p-H$_2$O $N=4-5$ is already enough due to the very strong interaction for the first and the much weaker interaction of the last two, respectively. 
Interestingly, for the deuterated species whose chemical potential lies in between the other curves, it takes between 10 and 15 endofullerene cages to reach a plateau of the chemical potential. Such a slow convergence is often an indicator of the critical region around a phase transition~\cite{iouchtchenko2018ground,serwatka2022ground}. 
The Schmidt gap $\Delta\lambda$ was recently found to be a good order parameter that signals the occurrence of a quantum phase transition for the free water chain~\cite{serwatka2023qpt}. It is presented in Fig.~\ref{fig:sites}b) and shows that for LiF, even a dimer has a Schmidt gap near zero, an indication that dipoles are already aligned. In contrast, HF, p-H$_2$O and DF (screened), o-D$_2$O (screened) have a constant value between 1 and 0.9 which means all these systems are in a disordered quantum phase where the rotational energy is dominant and prevents  dipolar alignment. The unscreened, deuterated species show a completely different behavior. For small chains, the systems are in a disordered phase but with increasing $N$, the Schmidt gap decreases. This means that for growing chains, the dipoles collectively align resulting in an ordered quantum phase. This slower formation of the dipole-ordered phase explains the slower convergence of the chemical potential. Based on the Schmidt gap, we can now identify three candidates that form a dipole-ordered quantum phase: screened LiF, and unscreened DF and o-D$_2$O. In these systems, it is possible to observe a quantum phase transition which makes them promising candidates as platform for quantum devices~\cite{latorre2004adiabatic,schutzhold2006adiabatic}.

Up to now, we only focused on chains that are symmetric with respect to a flip of all dipoles. However, in experimental realizations of these systems, this symmetry will always be broken by an anisotropic environment such as a carbon nanotube~\cite{biskupek2020bond}, a surface, or the fact that not all cages will be oriented identically. 
As shown in previous work, breaking the inversion symmetry splits the two-fold degenerate ground state of the ordered phase into two states with opposite polarization and a ferroelectric phase is formed~\cite{serwatka2023qpt,serwatka2022efield}. 
Such ferroelectric systems might be used as sensors or memory-devices~\cite{auciello1998physics,damjanovic2001ferroelectric,kim2018application}. 
For the three present species, we calculated the axial polarization for chains for which we applied a small axial electric field at the edge cages of the respective chains (see Fig.~\ref{fig:sites}c)). For the first two states, LiF exhibits a very high axial polarization $M_{z}$  of $\pm 0.84$, an indication of a ferroelectric phase. As previously observed, the polarization converges very quickly with $N$. 
Since the DF and o-D$_2$O chains are in the critical region, the convergence to a constant value is much slower, as already seen for the chemical potential. As shown in previous studies~\cite{serwatka2023qpt,serwatka2022efield}, the symmetry breaking leads to an energy gap of a few \SI{}{\per\centi\metre} between the two oppositely polarized states (see Fig.~\ref{fig:sites}d)). This means that it becomes possible to switch the polarization using a fundamental excitation. Hence, those ferroelectric systems might be used as dipolar molecular switches. As for the previous observables, for LiF,  the dimer already shows a bulk-like gap whereas DF and o-D$_2$O require longer chains to converge to a constant energy gap.

\begin{figure}[!htb]
\includegraphics[width=\columnwidth]{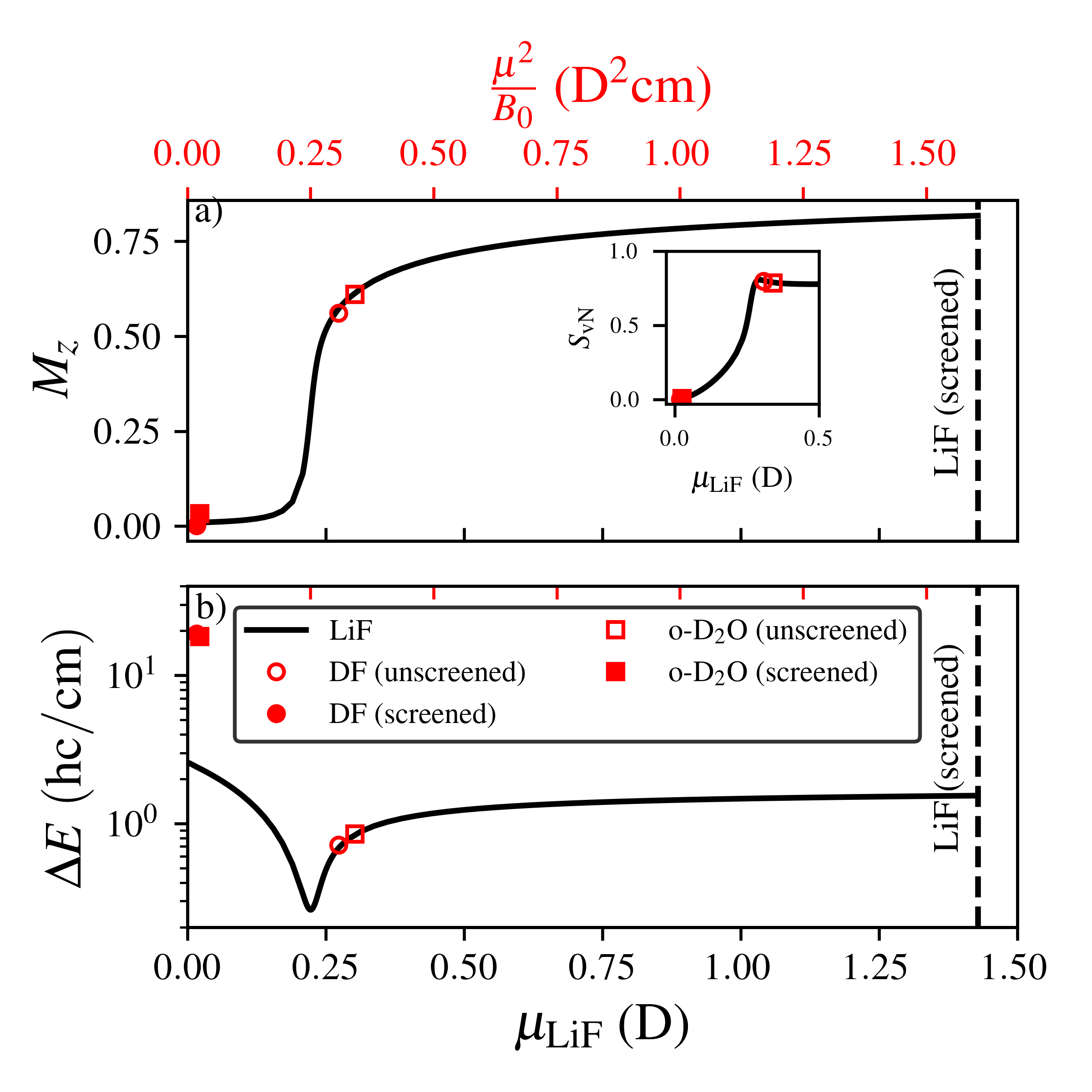}
\caption{a) Polarization and b) fundamental energy gap for LiF (black solid line), DF (red circles) and o-D$_2$O (red squares). All calculations are performed for $N=30$. An axial electric field $|\vec{\mu}\vec{F}|=\SI{0.5}{hc\per\centi\metre}$ is applied at the edge cages. For o-D$_2$O the rescaled dipolar interaction strength was calculated by using the arithmetic mean of $A_{e}$ and $C_{e}$. Inset: von-Neumann entanglement entropy for $N=30$ without electric fields.}
\label{fig:univbehaviour}
\end{figure}

\begin{figure*}[!htb]
\includegraphics[width=\textwidth]{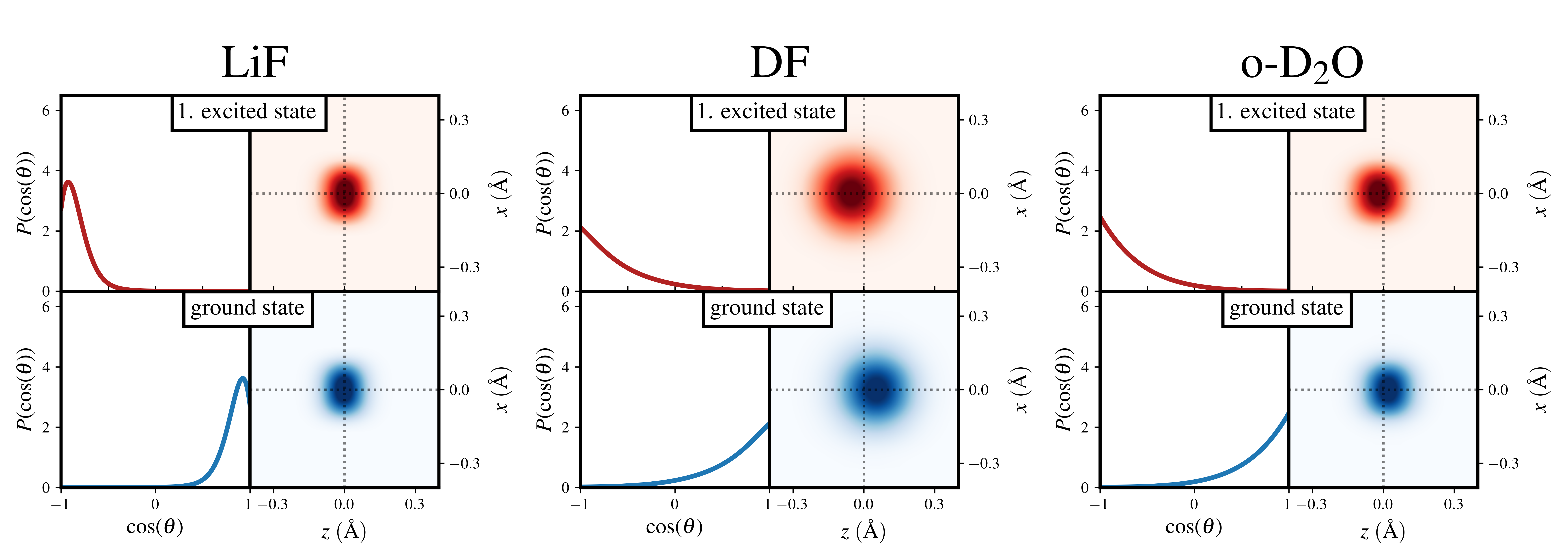}
\caption{Density distributions of LiF, DF and o-D$_{2}$O endofullerene chains calculated for the central site of chains with $N=29$. For every guest species, the angular density $P(\cos (\theta))$ (left figures) and the two-dimensional density (right panel) along the translational coordinates $x$ and $z$ are depicted.}
\label{fig:densities}
\end{figure*}

Although all three species form an dipole-ordered ferroelectric quantum phase and all of them show the same qualitative behavior with increasing system size, there is a quantitative difference between LiF, DF and o-D$_2$O whereas the last two are similar. Based on our initial statement, these differences should be caused by the diverging relative contributions of $\hat{T}_{\mathrm{rot}}$ and $\hat{V}_{\mathrm{dd}}$. 
To substantiate this assumption, we plot $S_{\mathrm{vN}}$, $M_{z}$ and $\Delta E$ for a LiF chain ($N=30)$ for different $\mu_{\mathrm{eff}}$ in Fig.~\ref{fig:univbehaviour}.  
The variable effective dipole moment is equivalent to the assumption of different screening effect whereas we assume similar LiF-cage interactions. Not very surprisingly, the dashed lines in Fig.~\ref{fig:univbehaviour} reveal that LiF with an effective dipole moment as found in C$_{60}$ is deep in the ordered phase. The quantum phase transition, marked by the maximum in $S_{\mathrm{vN}}$ or the minimum in $\Delta E$, occurs at a much smaller $\mu_{\mathrm{eff}}$ (below 0.25~D). In order to relate the respective properties of the DF and o-D$_2$ systems to the the LiF curves, we also plot their properties against $\frac{\mu_{\mathrm{eff}}^2}{B}$ shown on the red x-axis. This variable describes the dipole-interaction strength relative to the rotational energy term. What is apparent in Fig.~\ref{fig:univbehaviour} is the perfect match of the LiF curves and the points of the unscreened DF and o-D$_2$O chains. This proves a universal behavior of the three species that only depends on the relative strengths of $V_{\mathrm{dd}}$ and $T_{\mathrm{rot}}$. It also explains why DF and D$_2$O behave so similar because they have a similar $\frac{\mu^2_{\mathrm{eff}}}{B}$ ratio. This specific ratio positions them in the critical region near the phase transition which is in agreement with the slow convergence  we observed for various properties. 
The fact that these systems show a universal behavior that only depends on $\frac{\mu^2_{\mathrm{eff}}}{B}$ also means that the molecule-cage interactions $\hat{V}_{\mathrm{guest-cage}}$ and the translational term $\hat{T}_{\mathrm{trans}}$ are less important. That is also visible in the density distributions along $\cos (\theta)$ and along the center-of mass coordinates $x,z$ depicted in Fig.~\ref{fig:densities}. 
They look similar for all three systems. 
All show a strong polarization of opposite direction in the first two states. 
In all systems, the guest molecule is also strongly localized along the translational coordinates. 
LiF and o-D$_{2}$O are more strongly localized than DF due to their higher mass and larger van-der Waals radii.
Strong translational localization means that molecules behave like pinned rotors.
Hence, the translational term in Eq.~\eqref{eq:hamiltonian} is negligible. 
The fact that the peaks of the distributions for DF and o-D$_{2}$O are slightly off-center is due to the anisotropy of the molecule-cage interactions and the actual peak position depends on molecular orientation.

The universal behavior depends on the importance and the variance of $\hat{V}_{\mathrm{guest-cage}}$ and $\hat{T}_{\mathrm{trans}}$. 
For different cages where electrostatic or covalent interactions play a more important role and the interactions show a strong anisotropy, this universal behavior might not hold anymore because the terms Eq.~\eqref{eq:hamiltonian} are not negligible and they might differ strongly for different guest molecules. In this case a rescaling of $\hat{V}_{\mathrm{dd}}$ would not be possible. This might also be the case for higher excited states for which H$_{2}$O@C$_{60}$ and HF@C$_{60}$ are known to have a stronger rotation-translation coupling~\cite{krachmalnicoff2016dipolar,wespiser2022ro}. 
Likewise, in the disordered phase, there is no universal behavior because $\hat{V}_{\mathrm{dd}}$ is negligible and the system is thus solely determined by the total mass $M$, the rotational constants and $\hat{V}_{\mathrm{guest-cage}}$.

\begin{figure}[!htb]
\includegraphics[width=\columnwidth]{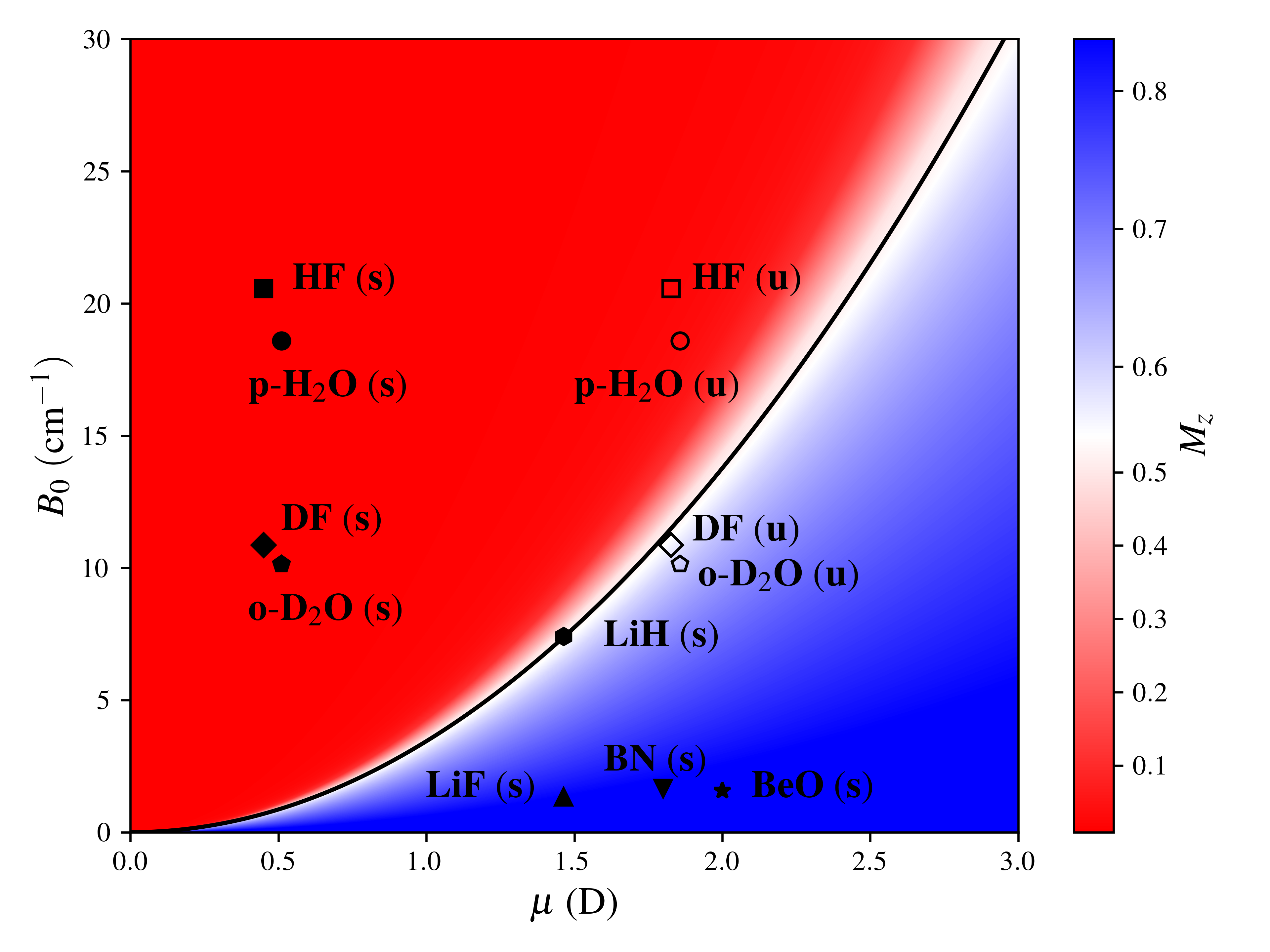}
\caption{Phase diagram for endofullerene chains with markers for various screened (s) and unscreened (us) species. The polarization is deduced from the polarization curve of LiF in Fig.~\ref{fig:univbehaviour} and the black curve is derived from the maximum of the entropy curve in Fig.~\ref{fig:univbehaviour}. The values for the screened dipole moments of LiH, BN and BeO are taken from \cite{dolgonos2014encapsulation}.}
\label{fig:phasediagram}
\end{figure}

But for the endofullerene systems studied here, the first two, oppositely polarized states show a universal behavior. Based on the results shown in Fig.~\ref{fig:univbehaviour} and by assuming minor contributions of $T_{\mathrm{trans}}$ and $V_{\mathrm{guest-cage}}$, we construct a phase diagram shown in Fig.~\ref{fig:phasediagram}. 
This depicts the occurrence of the two quantum phases (disordered and dipole-ordered), signalled by the order parameter $M_{z}$ in the symmetry-broken system, on the basis of $\mu_{\mathrm{eff}}$ and $B$. For the aforementioned assumption, we predict other possible molecules that can form dipole-ordered, ferroelectric chains. Fig.~\ref{fig:phasediagram} underlines that it is always a trade-off between rotational constants and effective electric dipole moment. The optimal case to construct a chain of endofullerenes are molecules with small rotational constants and large effective dipole moment. For a fixed molecular species, the phase, and thus the ferroelectricity of the chain can be modified by isotopic substitution which changes $B$ and shifts the points in Fig.~\ref{fig:phasediagram} along the y-axis or by modifying the cage, which changes $\mu_{\mathrm{eff}}$ and shifts the points along the x-axis. According to Fig.~\ref{fig:phasediagram}, strongly polar diatomics such as BN or BeO inserted in C$_{60}$ are very promising candidates for ferroelectric systems that do not even require further modifications of isotopologues or cages. Of great interest are also systems such as LiH@C$_{60}$ that are near the phase transition, a region with a rich variety of physical phenomena~\cite{coleman2005quantum,sachdev2011quantumcrit}.

Many-body endohedral and other cage systems show a great potential for applications as ferroelectric material or as quantum devices. 
Our DMRG approach provides the tool to directly include the atom-specific interactions without relying on a simplified model. Future work will incorporate the study of different guest-cage systems, especially substituted fullerenes and heterofullerene cages and their impact on the quantum phases. An appealing feature of the endohedral systems is their modularity since it is possible to build systems with different dimensionality and symmetry. In these systems, the number of neighboring molecules will differ which might lead to entangled systems even for strongly screened guest molecules like H$_2$O@C$_{60}$.
Interesting electronic effects such as vibronic couplings will be explored in future work \cite{hauser2022vibronic}.
Moreover, the extension to finite temperatures is a necessary step on the way towards an understanding the nature of actual physical realizations of these systems.

\section{Acknowledgments}
This research was supported by the Natural Sciences and Engineering Research Council of Canada (NSERC), the Ontario Ministry of Research and Innovation (MRI), the Canada Research Chair program (950-231024), the Canada Foundation for Innovation (CFI) (project No. 35232) and Compute Canada. T. S. acknowledges a Walter-Benjamin funding of the Deutsche Forschungsgemeinschaft (Projektnummer 503971734). 

\section{Supporting Information}
Details of the translation-rotation basis, dipole-dipole interaction matrix elements, molecular parameters, and of the proposed molecular-cage interaction potential for LiF@C$_{60}$ are presented.
This material is available free of charge via the Internet at http://pubs.acs.org.


\begin{thebibliography}{40}%
\makeatletter
\providecommand \@ifxundefined [1]{%
 \@ifx{#1\undefined}
}%
\providecommand \@ifnum [1]{%
 \ifnum #1\expandafter \@firstoftwo
 \else \expandafter \@secondoftwo
 \fi
}%
\providecommand \@ifx [1]{%
 \ifx #1\expandafter \@firstoftwo
 \else \expandafter \@secondoftwo
 \fi
}%
\providecommand \natexlab [1]{#1}%
\providecommand \enquote  [1]{``#1''}%
\providecommand \bibnamefont  [1]{#1}%
\providecommand \bibfnamefont [1]{#1}%
\providecommand \citenamefont [1]{#1}%
\providecommand \href@noop [0]{\@secondoftwo}%
\providecommand \href [0]{\begingroup \@sanitize@url \@href}%
\providecommand \@href[1]{\@@startlink{#1}\@@href}%
\providecommand \@@href[1]{\endgroup#1\@@endlink}%
\providecommand \@sanitize@url [0]{\catcode `\\12\catcode `\$12\catcode
  `\&12\catcode `\#12\catcode `\^12\catcode `\_12\catcode `\%12\relax}%
\providecommand \@@startlink[1]{}%
\providecommand \@@endlink[0]{}%
\providecommand \url  [0]{\begingroup\@sanitize@url \@url }%
\providecommand \@url [1]{\endgroup\@href {#1}{\urlprefix }}%
\providecommand \urlprefix  [0]{URL }%
\providecommand \Eprint [0]{\href }%
\providecommand \doibase [0]{http://dx.doi.org/}%
\providecommand \selectlanguage [0]{\@gobble}%
\providecommand \bibinfo  [0]{\@secondoftwo}%
\providecommand \bibfield  [0]{\@secondoftwo}%
\providecommand \translation [1]{[#1]}%
\providecommand \BibitemOpen [0]{}%
\providecommand \bibitemStop [0]{}%
\providecommand \bibitemNoStop [0]{.\EOS\space}%
\providecommand \EOS [0]{\spacefactor3000\relax}%
\providecommand \BibitemShut  [1]{\csname bibitem#1\endcsname}%
\let\auto@bib@innerbib\@empty
\bibitem [{\citenamefont {Murata}\ \emph {et~al.}(2008)\citenamefont {Murata},
  \citenamefont {Murata},\ and\ \citenamefont {Komatsu}}]{murata2008surgery}%
  \BibitemOpen
  \bibfield  {author} {\bibinfo {author} {\bibfnamefont {Michihisa}\
  \bibnamefont {Murata}}, \bibinfo {author} {\bibfnamefont {Yasujiro}\
  \bibnamefont {Murata}}, \ and\ \bibinfo {author} {\bibfnamefont {Koichi}\
  \bibnamefont {Komatsu}},\ }\bibfield  {title} {\enquote {\bibinfo {title}
  {Surgery of fullerenes},}\ }\href@noop {} {\bibfield  {journal} {\bibinfo
  {journal} {Chem. Commun.}\ ,\ \bibinfo {pages} {6083--6094}} (\bibinfo {year}
  {2008})}\BibitemShut {NoStop}%
\bibitem [{\citenamefont {Kurotobi}\ and\ \citenamefont
  {Murata}(2011)}]{kurotobi2011single}%
  \BibitemOpen
  \bibfield  {author} {\bibinfo {author} {\bibfnamefont {Kei}\ \bibnamefont
  {Kurotobi}}\ and\ \bibinfo {author} {\bibfnamefont {Yasujiro}\ \bibnamefont
  {Murata}},\ }\bibfield  {title} {\enquote {\bibinfo {title} {A single
  molecule of water encapsulated in fullerene c60},}\ }\href@noop {} {\bibfield
   {journal} {\bibinfo  {journal} {Science}\ }\textbf {\bibinfo {volume}
  {333}},\ \bibinfo {pages} {613--616} (\bibinfo {year} {2011})}\BibitemShut
  {NoStop}%
\bibitem [{\citenamefont {Bloodworth}\ and\ \citenamefont
  {Whitby}(2022)}]{bloodworth2022synthesis}%
  \BibitemOpen
  \bibfield  {author} {\bibinfo {author} {\bibfnamefont {Sally}\ \bibnamefont
  {Bloodworth}}\ and\ \bibinfo {author} {\bibfnamefont {Richard~J}\
  \bibnamefont {Whitby}},\ }\bibfield  {title} {\enquote {\bibinfo {title}
  {Synthesis of endohedral fullerenes by molecular surgery},}\ }\href@noop {}
  {\bibfield  {journal} {\bibinfo  {journal} {Commun. Chem.}\ }\textbf
  {\bibinfo {volume} {5}},\ \bibinfo {pages} {1--14} (\bibinfo {year}
  {2022})}\BibitemShut {NoStop}%
\bibitem [{\citenamefont {Kohama}\ \emph {et~al.}(2009)\citenamefont {Kohama},
  \citenamefont {Rachi}, \citenamefont {Jing}, \citenamefont {Li},
  \citenamefont {Tang}, \citenamefont {Kumashiro}, \citenamefont {Izumisawa},
  \citenamefont {Kawaji}, \citenamefont {Atake}, \citenamefont {Sawa} \emph
  {et~al.}}]{kohama2009rotational}%
  \BibitemOpen
  \bibfield  {author} {\bibinfo {author} {\bibfnamefont {Yoshimitsu}\
  \bibnamefont {Kohama}}, \bibinfo {author} {\bibfnamefont {Takeshi}\
  \bibnamefont {Rachi}}, \bibinfo {author} {\bibfnamefont {Ju}~\bibnamefont
  {Jing}}, \bibinfo {author} {\bibfnamefont {Zhaofei}\ \bibnamefont {Li}},
  \bibinfo {author} {\bibfnamefont {Jun}\ \bibnamefont {Tang}}, \bibinfo
  {author} {\bibfnamefont {Ryotaro}\ \bibnamefont {Kumashiro}}, \bibinfo
  {author} {\bibfnamefont {Satoru}\ \bibnamefont {Izumisawa}}, \bibinfo
  {author} {\bibfnamefont {Hitoshi}\ \bibnamefont {Kawaji}}, \bibinfo {author}
  {\bibfnamefont {Tooru}\ \bibnamefont {Atake}}, \bibinfo {author}
  {\bibfnamefont {Hiroshi}\ \bibnamefont {Sawa}},  \emph {et~al.},\ }\bibfield
  {title} {\enquote {\bibinfo {title} {Rotational sublevels of an
  ortho-hydrogen molecule encapsulated in an isotropic c 60 cage},}\
  }\href@noop {} {\bibfield  {journal} {\bibinfo  {journal} {Phys. Rev. Lett.}\
  }\textbf {\bibinfo {volume} {103}},\ \bibinfo {pages} {073001} (\bibinfo
  {year} {2009})}\BibitemShut {NoStop}%
\bibitem [{\citenamefont {Suzuki}\ \emph {et~al.}(2019)\citenamefont {Suzuki},
  \citenamefont {Nakano}, \citenamefont {Hashikawa},\ and\ \citenamefont
  {Murata}}]{suzuki2019rotational}%
  \BibitemOpen
  \bibfield  {author} {\bibinfo {author} {\bibfnamefont {Hal}\ \bibnamefont
  {Suzuki}}, \bibinfo {author} {\bibfnamefont {Motohiro}\ \bibnamefont
  {Nakano}}, \bibinfo {author} {\bibfnamefont {Yoshifumi}\ \bibnamefont
  {Hashikawa}}, \ and\ \bibinfo {author} {\bibfnamefont {Yasujiro}\
  \bibnamefont {Murata}},\ }\bibfield  {title} {\enquote {\bibinfo {title}
  {Rotational motion and nuclear spin interconversion of h2o encapsulated in
  c60 appearing in the low-temperature heat capacity},}\ }\href@noop {}
  {\bibfield  {journal} {\bibinfo  {journal} {J. Phys. Chem. Lett.}\ }\textbf
  {\bibinfo {volume} {10}},\ \bibinfo {pages} {1306--1311} (\bibinfo {year}
  {2019})}\BibitemShut {NoStop}%
\bibitem [{\citenamefont {Beduz}\ \emph {et~al.}(2012)\citenamefont {Beduz},
  \citenamefont {Carravetta}, \citenamefont {Chen}, \citenamefont
  {Concistr{\`e}}, \citenamefont {Denning}, \citenamefont {Frunzi},
  \citenamefont {Horsewill}, \citenamefont {Johannessen}, \citenamefont
  {Lawler}, \citenamefont {Lei} \emph {et~al.}}]{beduz2012quantum}%
  \BibitemOpen
  \bibfield  {author} {\bibinfo {author} {\bibfnamefont {Carlo}\ \bibnamefont
  {Beduz}}, \bibinfo {author} {\bibfnamefont {Marina}\ \bibnamefont
  {Carravetta}}, \bibinfo {author} {\bibfnamefont {Judy Y-C}\ \bibnamefont
  {Chen}}, \bibinfo {author} {\bibfnamefont {Maria}\ \bibnamefont
  {Concistr{\`e}}}, \bibinfo {author} {\bibfnamefont {Mark}\ \bibnamefont
  {Denning}}, \bibinfo {author} {\bibfnamefont {Michael}\ \bibnamefont
  {Frunzi}}, \bibinfo {author} {\bibfnamefont {Anthony~J}\ \bibnamefont
  {Horsewill}}, \bibinfo {author} {\bibfnamefont {Ole~G}\ \bibnamefont
  {Johannessen}}, \bibinfo {author} {\bibfnamefont {Ronald}\ \bibnamefont
  {Lawler}}, \bibinfo {author} {\bibfnamefont {Xuegong}\ \bibnamefont {Lei}},
  \emph {et~al.},\ }\bibfield  {title} {\enquote {\bibinfo {title} {Quantum
  rotation of ortho and para-water encapsulated in a fullerene cage},}\
  }\href@noop {} {\bibfield  {journal} {\bibinfo  {journal} {Proc. Natl. Acad.
  Sci. U.S.A.}\ }\textbf {\bibinfo {volume} {109}},\ \bibinfo {pages}
  {12894--12898} (\bibinfo {year} {2012})}\BibitemShut {NoStop}%
\bibitem [{\citenamefont {Mamone}\ \emph {et~al.}(2014)\citenamefont {Mamone},
  \citenamefont {Concistr{\`e}}, \citenamefont {Carignani}, \citenamefont
  {Meier}, \citenamefont {Krachmalnicoff}, \citenamefont {Johannessen},
  \citenamefont {Lei}, \citenamefont {Li}, \citenamefont {Denning},
  \citenamefont {Carravetta} \emph {et~al.}}]{mamone2014nuclear}%
  \BibitemOpen
  \bibfield  {author} {\bibinfo {author} {\bibfnamefont {Salvatore}\
  \bibnamefont {Mamone}}, \bibinfo {author} {\bibfnamefont {Maria}\
  \bibnamefont {Concistr{\`e}}}, \bibinfo {author} {\bibfnamefont {Elisa}\
  \bibnamefont {Carignani}}, \bibinfo {author} {\bibfnamefont {Benno}\
  \bibnamefont {Meier}}, \bibinfo {author} {\bibfnamefont {Andrea}\
  \bibnamefont {Krachmalnicoff}}, \bibinfo {author} {\bibfnamefont {Ole~G}\
  \bibnamefont {Johannessen}}, \bibinfo {author} {\bibfnamefont {Xuegong}\
  \bibnamefont {Lei}}, \bibinfo {author} {\bibfnamefont {Yongjun}\ \bibnamefont
  {Li}}, \bibinfo {author} {\bibfnamefont {Mark}\ \bibnamefont {Denning}},
  \bibinfo {author} {\bibfnamefont {Marina}\ \bibnamefont {Carravetta}},  \emph
  {et~al.},\ }\bibfield  {title} {\enquote {\bibinfo {title} {Nuclear spin
  conversion of water inside fullerene cages detected by low-temperature
  nuclear magnetic resonance},}\ }\href@noop {} {\bibfield  {journal} {\bibinfo
   {journal} {J. Chem. Phys.}\ }\textbf {\bibinfo {volume} {140}},\ \bibinfo
  {pages} {194306} (\bibinfo {year} {2014})}\BibitemShut {NoStop}%
\bibitem [{\citenamefont {Yang}\ \emph {et~al.}(2010)\citenamefont {Yang},
  \citenamefont {Xu}, \citenamefont {Wei}, \citenamefont {Feng},\ and\
  \citenamefont {Suter}}]{yang2010quantum}%
  \BibitemOpen
  \bibfield  {author} {\bibinfo {author} {\bibfnamefont {Wan~Li}\ \bibnamefont
  {Yang}}, \bibinfo {author} {\bibfnamefont {Zhen~Yu}\ \bibnamefont {Xu}},
  \bibinfo {author} {\bibfnamefont {Hua}\ \bibnamefont {Wei}}, \bibinfo
  {author} {\bibfnamefont {Mang}\ \bibnamefont {Feng}}, \ and\ \bibinfo
  {author} {\bibfnamefont {Dieter}\ \bibnamefont {Suter}},\ }\bibfield  {title}
  {\enquote {\bibinfo {title} {Quantum-information-processing architecture with
  endohedral fullerenes in a carbon nanotube},}\ }\href@noop {} {\bibfield
  {journal} {\bibinfo  {journal} {Phys. Rev. A}\ }\textbf {\bibinfo {volume}
  {81}},\ \bibinfo {pages} {032303} (\bibinfo {year} {2010})}\BibitemShut
  {NoStop}%
\bibitem [{\citenamefont {Zhou}\ \emph {et~al.}(2022)\citenamefont {Zhou},
  \citenamefont {Yuan}, \citenamefont {Wang}, \citenamefont {Ling},
  \citenamefont {Fu}, \citenamefont {Fang}, \citenamefont {Wang}, \citenamefont
  {Liu}, \citenamefont {Porfyrakis}, \citenamefont {Briggs} \emph
  {et~al.}}]{zhou2022implementation}%
  \BibitemOpen
  \bibfield  {author} {\bibinfo {author} {\bibfnamefont {Shen}\ \bibnamefont
  {Zhou}}, \bibinfo {author} {\bibfnamefont {Jiayue}\ \bibnamefont {Yuan}},
  \bibinfo {author} {\bibfnamefont {Zi-Yu}\ \bibnamefont {Wang}}, \bibinfo
  {author} {\bibfnamefont {Kun}\ \bibnamefont {Ling}}, \bibinfo {author}
  {\bibfnamefont {Peng-Xiang}\ \bibnamefont {Fu}}, \bibinfo {author}
  {\bibfnamefont {Yu-Hui}\ \bibnamefont {Fang}}, \bibinfo {author}
  {\bibfnamefont {Ye-Xin}\ \bibnamefont {Wang}}, \bibinfo {author}
  {\bibfnamefont {Zheng}\ \bibnamefont {Liu}}, \bibinfo {author} {\bibfnamefont
  {Kyriakos}\ \bibnamefont {Porfyrakis}}, \bibinfo {author} {\bibfnamefont
  {G~Andrew~D}\ \bibnamefont {Briggs}},  \emph {et~al.},\ }\bibfield  {title}
  {\enquote {\bibinfo {title} {Implementation of quantum level addressability
  and geometric phase manipulation in aligned endohedral fullerene qudits},}\
  }\href@noop {} {\bibfield  {journal} {\bibinfo  {journal} {Angew. Chem.}\
  }\textbf {\bibinfo {volume} {134}},\ \bibinfo {pages} {e202115263} (\bibinfo
  {year} {2022})}\BibitemShut {NoStop}%
\bibitem [{\citenamefont {David}\ \emph {et~al.}(1991)\citenamefont {David},
  \citenamefont {Ibberson}, \citenamefont {Matthewman}, \citenamefont
  {Prassides}, \citenamefont {Dennis}, \citenamefont {Hare}, \citenamefont
  {Kroto}, \citenamefont {Taylor},\ and\ \citenamefont
  {Walton}}]{david1991crystal}%
  \BibitemOpen
  \bibfield  {author} {\bibinfo {author} {\bibfnamefont {William~IF}\
  \bibnamefont {David}}, \bibinfo {author} {\bibfnamefont {Richard~M}\
  \bibnamefont {Ibberson}}, \bibinfo {author} {\bibfnamefont {Judy~C}\
  \bibnamefont {Matthewman}}, \bibinfo {author} {\bibfnamefont {Kosmas}\
  \bibnamefont {Prassides}}, \bibinfo {author} {\bibfnamefont {T~John~S}\
  \bibnamefont {Dennis}}, \bibinfo {author} {\bibfnamefont {Jonathan~P}\
  \bibnamefont {Hare}}, \bibinfo {author} {\bibfnamefont {Harold~W}\
  \bibnamefont {Kroto}}, \bibinfo {author} {\bibfnamefont {Roger}\ \bibnamefont
  {Taylor}}, \ and\ \bibinfo {author} {\bibfnamefont {David~RM}\ \bibnamefont
  {Walton}},\ }\bibfield  {title} {\enquote {\bibinfo {title} {Crystal
  structure and bonding of ordered c60},}\ }\href@noop {} {\bibfield  {journal}
  {\bibinfo  {journal} {Nature}\ }\textbf {\bibinfo {volume} {353}},\ \bibinfo
  {pages} {147--149} (\bibinfo {year} {1991})}\BibitemShut {NoStop}%
\bibitem [{\citenamefont {Shi}\ \emph {et~al.}(1994)\citenamefont {Shi},
  \citenamefont {Caldwell}, \citenamefont {Chen},\ and\ \citenamefont
  {Mirkin}}]{shi1994well}%
  \BibitemOpen
  \bibfield  {author} {\bibinfo {author} {\bibfnamefont {Xiaobo}\ \bibnamefont
  {Shi}}, \bibinfo {author} {\bibfnamefont {W~Brett}\ \bibnamefont {Caldwell}},
  \bibinfo {author} {\bibfnamefont {Kaimin}\ \bibnamefont {Chen}}, \ and\
  \bibinfo {author} {\bibfnamefont {Chad~A}\ \bibnamefont {Mirkin}},\
  }\bibfield  {title} {\enquote {\bibinfo {title} {A well-defined
  surface-confinable fullerene: monolayer self-assembly on au (111)},}\
  }\href@noop {} {\bibfield  {journal} {\bibinfo  {journal} {J. Am. Chem.
  Soc.}\ }\textbf {\bibinfo {volume} {116}},\ \bibinfo {pages} {11598--11599}
  (\bibinfo {year} {1994})}\BibitemShut {NoStop}%
\bibitem [{\citenamefont {Gulino}\ \emph {et~al.}(2005)\citenamefont {Gulino},
  \citenamefont {Bazzano}, \citenamefont {Condorelli}, \citenamefont
  {Giuffrida}, \citenamefont {Mineo}, \citenamefont {Satriano}, \citenamefont
  {Scamporrino}, \citenamefont {Ventimiglia}, \citenamefont {Vitalini},\ and\
  \citenamefont {Fragala}}]{gulino2005engineered}%
  \BibitemOpen
  \bibfield  {author} {\bibinfo {author} {\bibfnamefont {Antonino}\
  \bibnamefont {Gulino}}, \bibinfo {author} {\bibfnamefont {Sebastiano}\
  \bibnamefont {Bazzano}}, \bibinfo {author} {\bibfnamefont {Guglielmo~G}\
  \bibnamefont {Condorelli}}, \bibinfo {author} {\bibfnamefont {Salvatore}\
  \bibnamefont {Giuffrida}}, \bibinfo {author} {\bibfnamefont {Placido}\
  \bibnamefont {Mineo}}, \bibinfo {author} {\bibfnamefont {Cristina}\
  \bibnamefont {Satriano}}, \bibinfo {author} {\bibfnamefont {Emilio}\
  \bibnamefont {Scamporrino}}, \bibinfo {author} {\bibfnamefont {Giorgio}\
  \bibnamefont {Ventimiglia}}, \bibinfo {author} {\bibfnamefont {Daniele}\
  \bibnamefont {Vitalini}}, \ and\ \bibinfo {author} {\bibfnamefont {Ignazio}\
  \bibnamefont {Fragala}},\ }\bibfield  {title} {\enquote {\bibinfo {title}
  {Engineered silica surfaces with an assembled c60 fullerene monolayer},}\
  }\href@noop {} {\bibfield  {journal} {\bibinfo  {journal} {Chem. Mater.}\
  }\textbf {\bibinfo {volume} {17}},\ \bibinfo {pages} {1079--1084} (\bibinfo
  {year} {2005})}\BibitemShut {NoStop}%
\bibitem [{\citenamefont {Hou}\ \emph {et~al.}(2022)\citenamefont {Hou},
  \citenamefont {Cui}, \citenamefont {Guan}, \citenamefont {Wang},
  \citenamefont {Li}, \citenamefont {Liu}, \citenamefont {Zhu},\ and\
  \citenamefont {Zheng}}]{hou2022synthesis}%
  \BibitemOpen
  \bibfield  {author} {\bibinfo {author} {\bibfnamefont {Lingxiang}\
  \bibnamefont {Hou}}, \bibinfo {author} {\bibfnamefont {Xueping}\ \bibnamefont
  {Cui}}, \bibinfo {author} {\bibfnamefont {Bo}~\bibnamefont {Guan}}, \bibinfo
  {author} {\bibfnamefont {Shaozhi}\ \bibnamefont {Wang}}, \bibinfo {author}
  {\bibfnamefont {Ruian}\ \bibnamefont {Li}}, \bibinfo {author} {\bibfnamefont
  {Yunqi}\ \bibnamefont {Liu}}, \bibinfo {author} {\bibfnamefont {Daoben}\
  \bibnamefont {Zhu}}, \ and\ \bibinfo {author} {\bibfnamefont {Jian}\
  \bibnamefont {Zheng}},\ }\bibfield  {title} {\enquote {\bibinfo {title}
  {Synthesis of a monolayer fullerene network},}\ }\href@noop {} {\bibfield
  {journal} {\bibinfo  {journal} {Nature}\ }\textbf {\bibinfo {volume} {606}},\
  \bibinfo {pages} {507--510} (\bibinfo {year} {2022})}\BibitemShut {NoStop}%
\bibitem [{\citenamefont {Biskupek}\ \emph {et~al.}(2020)\citenamefont
  {Biskupek}, \citenamefont {Skowron}, \citenamefont {Stoppiello},
  \citenamefont {Rance}, \citenamefont {Alom}, \citenamefont {Fung},
  \citenamefont {Whitby}, \citenamefont {Levitt}, \citenamefont {Ramasse},
  \citenamefont {Kaiser} \emph {et~al.}}]{biskupek2020bond}%
  \BibitemOpen
  \bibfield  {author} {\bibinfo {author} {\bibfnamefont {Johannes}\
  \bibnamefont {Biskupek}}, \bibinfo {author} {\bibfnamefont {Stephen~T}\
  \bibnamefont {Skowron}}, \bibinfo {author} {\bibfnamefont {Craig~T}\
  \bibnamefont {Stoppiello}}, \bibinfo {author} {\bibfnamefont {Graham~A}\
  \bibnamefont {Rance}}, \bibinfo {author} {\bibfnamefont {Shamim}\
  \bibnamefont {Alom}}, \bibinfo {author} {\bibfnamefont {Kayleigh~LY}\
  \bibnamefont {Fung}}, \bibinfo {author} {\bibfnamefont {Richard~J}\
  \bibnamefont {Whitby}}, \bibinfo {author} {\bibfnamefont {Malcolm~H}\
  \bibnamefont {Levitt}}, \bibinfo {author} {\bibfnamefont {Quentin~M}\
  \bibnamefont {Ramasse}}, \bibinfo {author} {\bibfnamefont {Ute}\ \bibnamefont
  {Kaiser}},  \emph {et~al.},\ }\bibfield  {title} {\enquote {\bibinfo {title}
  {Bond dissociation and reactivity of hf and h2o in a nano test tube},}\
  }\href@noop {} {\bibfield  {journal} {\bibinfo  {journal} {ACS Nano}\
  }\textbf {\bibinfo {volume} {14}},\ \bibinfo {pages} {11178--11189} (\bibinfo
  {year} {2020})}\BibitemShut {NoStop}%
\bibitem [{\citenamefont {Cioslowski}\ and\ \citenamefont
  {Nanayakkara}(1992)}]{cioslowski1992endohedral}%
  \BibitemOpen
  \bibfield  {author} {\bibinfo {author} {\bibfnamefont {Jerzy}\ \bibnamefont
  {Cioslowski}}\ and\ \bibinfo {author} {\bibfnamefont {Asiri}\ \bibnamefont
  {Nanayakkara}},\ }\bibfield  {title} {\enquote {\bibinfo {title} {Endohedral
  fullerites: a new class of ferroelectric materials},}\ }\href@noop {}
  {\bibfield  {journal} {\bibinfo  {journal} {Phys. Rev. Lett.}\ }\textbf
  {\bibinfo {volume} {69}},\ \bibinfo {pages} {2871} (\bibinfo {year}
  {1992})}\BibitemShut {NoStop}%
\bibitem [{\citenamefont {Moses}\ \emph {et~al.}(2015)\citenamefont {Moses},
  \citenamefont {Covey}, \citenamefont {Miecnikowski}, \citenamefont {Yan},
  \citenamefont {Gadway}, \citenamefont {Ye},\ and\ \citenamefont
  {Jin}}]{opticallattice2015}%
  \BibitemOpen
  \bibfield  {author} {\bibinfo {author} {\bibfnamefont {Steven~A.}\
  \bibnamefont {Moses}}, \bibinfo {author} {\bibfnamefont {Jacob~P.}\
  \bibnamefont {Covey}}, \bibinfo {author} {\bibfnamefont {Matthew~T.}\
  \bibnamefont {Miecnikowski}}, \bibinfo {author} {\bibfnamefont
  {Bo}~\bibnamefont {Yan}}, \bibinfo {author} {\bibfnamefont {Bryce}\
  \bibnamefont {Gadway}}, \bibinfo {author} {\bibfnamefont {Jun}\ \bibnamefont
  {Ye}}, \ and\ \bibinfo {author} {\bibfnamefont {Deborah~S.}\ \bibnamefont
  {Jin}},\ }\bibfield  {title} {\enquote {\bibinfo {title} {Creation of a
  low-entropy quantum gas of polar molecules in an optical lattice},}\ }\href
  {\doibase 10.1126/science.aac6400} {\bibfield  {journal} {\bibinfo  {journal}
  {Science}\ }\textbf {\bibinfo {volume} {350}},\ \bibinfo {pages} {659--662}
  (\bibinfo {year} {2015})}\BibitemShut {NoStop}%
\bibitem [{\citenamefont {White}(1992)}]{white1992density}%
  \BibitemOpen
  \bibfield  {author} {\bibinfo {author} {\bibfnamefont {Steven~R}\
  \bibnamefont {White}},\ }\bibfield  {title} {\enquote {\bibinfo {title}
  {Density matrix formulation for quantum renormalization groups},}\
  }\href@noop {} {\bibfield  {journal} {\bibinfo  {journal} {Phys. Rev. Lett}\
  }\textbf {\bibinfo {volume} {69}},\ \bibinfo {pages} {2863} (\bibinfo {year}
  {1992})}\BibitemShut {NoStop}%
\bibitem [{\citenamefont {Fishman}\ \emph {et~al.}(2022)\citenamefont
  {Fishman}, \citenamefont {White},\ and\ \citenamefont
  {Stoudenmire}}]{fishman2022itensor}%
  \BibitemOpen
  \bibfield  {author} {\bibinfo {author} {\bibfnamefont {Matthew}\ \bibnamefont
  {Fishman}}, \bibinfo {author} {\bibfnamefont {Steven}\ \bibnamefont {White}},
  \ and\ \bibinfo {author} {\bibfnamefont {Edwin}\ \bibnamefont
  {Stoudenmire}},\ }\bibfield  {title} {\enquote {\bibinfo {title} {The itensor
  software library for tensor network calculations},}\ }\href@noop {}
  {\bibfield  {journal} {\bibinfo  {journal} {SciPost Physics Codebases}\ ,\
  \bibinfo {pages} {004}} (\bibinfo {year} {2022})}\BibitemShut {NoStop}%
\bibitem [{\citenamefont {Serwatka}\ and\ \citenamefont
  {Roy}(2022{\natexlab{a}})}]{serwatka2022ground}%
  \BibitemOpen
  \bibfield  {author} {\bibinfo {author} {\bibfnamefont {Tobias}\ \bibnamefont
  {Serwatka}}\ and\ \bibinfo {author} {\bibfnamefont {Pierre-Nicholas}\
  \bibnamefont {Roy}},\ }\bibfield  {title} {\enquote {\bibinfo {title} {Ground
  state of asymmetric tops with dmrg: Water in one dimension},}\ }\href@noop {}
  {\bibfield  {journal} {\bibinfo  {journal} {J. Chem. Phys.}\ }\textbf
  {\bibinfo {volume} {156}},\ \bibinfo {pages} {044116} (\bibinfo {year}
  {2022}{\natexlab{a}})}\BibitemShut {NoStop}%
\bibitem [{\citenamefont {Serwatka}\ and\ \citenamefont
  {Roy}(2022{\natexlab{b}})}]{serwatka2022efield}%
  \BibitemOpen
  \bibfield  {author} {\bibinfo {author} {\bibfnamefont {Tobias}\ \bibnamefont
  {Serwatka}}\ and\ \bibinfo {author} {\bibfnamefont {Pierre-Nicholas}\
  \bibnamefont {Roy}},\ }\bibfield  {title} {\enquote {\bibinfo {title}
  {Ferroelectric water chains in carbon nanotubes: Creation and manipulation of
  ordered quantum phases},}\ }\href@noop {} {\bibfield  {journal} {\bibinfo
  {journal} {J. Chem. Phys.}\ }\textbf {\bibinfo {volume} {157}},\ \bibinfo
  {pages} {234301} (\bibinfo {year} {2022}{\natexlab{b}})}\BibitemShut
  {NoStop}%
\bibitem [{\citenamefont {Serwatka}\ \emph {et~al.}(2023)\citenamefont
  {Serwatka}, \citenamefont {Melko}, \citenamefont {Burkov},\ and\
  \citenamefont {Roy}}]{serwatka2023qpt}%
  \BibitemOpen
  \bibfield  {author} {\bibinfo {author} {\bibfnamefont {T.}~\bibnamefont
  {Serwatka}}, \bibinfo {author} {\bibfnamefont {R.~G.}\ \bibnamefont {Melko}},
  \bibinfo {author} {\bibfnamefont {A.}~\bibnamefont {Burkov}}, \ and\ \bibinfo
  {author} {\bibfnamefont {P.-N.}\ \bibnamefont {Roy}},\ }\bibfield  {title}
  {\enquote {\bibinfo {title} {Quantum phase transition in the one-dimensional
  water chain},}\ }\href {\doibase 10.1103/PhysRevLett.130.026201} {\bibfield
  {journal} {\bibinfo  {journal} {Phys. Rev. Lett.}\ }\textbf {\bibinfo
  {volume} {130}},\ \bibinfo {pages} {026201} (\bibinfo {year}
  {2023})}\BibitemShut {NoStop}%
\bibitem [{\citenamefont {Meier}\ \emph {et~al.}(2015)\citenamefont {Meier},
  \citenamefont {Mamone}, \citenamefont {Concistr{\`e}}, \citenamefont
  {Alonso-Valdesueiro}, \citenamefont {Krachmalnicoff}, \citenamefont
  {Whitby},\ and\ \citenamefont {Levitt}}]{meier2015electrical}%
  \BibitemOpen
  \bibfield  {author} {\bibinfo {author} {\bibfnamefont {Benno}\ \bibnamefont
  {Meier}}, \bibinfo {author} {\bibfnamefont {Salvatore}\ \bibnamefont
  {Mamone}}, \bibinfo {author} {\bibfnamefont {Maria}\ \bibnamefont
  {Concistr{\`e}}}, \bibinfo {author} {\bibfnamefont {Javier}\ \bibnamefont
  {Alonso-Valdesueiro}}, \bibinfo {author} {\bibfnamefont {Andrea}\
  \bibnamefont {Krachmalnicoff}}, \bibinfo {author} {\bibfnamefont {Richard~J}\
  \bibnamefont {Whitby}}, \ and\ \bibinfo {author} {\bibfnamefont {Malcolm~H}\
  \bibnamefont {Levitt}},\ }\bibfield  {title} {\enquote {\bibinfo {title}
  {Electrical detection of ortho--para conversion in fullerene-encapsulated
  water},}\ }\href@noop {} {\bibfield  {journal} {\bibinfo  {journal} {Nat.
  Commun.}\ }\textbf {\bibinfo {volume} {6}},\ \bibinfo {pages} {1--4}
  (\bibinfo {year} {2015})}\BibitemShut {NoStop}%
\bibitem [{\citenamefont {Silvestrelli}\ \emph {et~al.}(2022)\citenamefont
  {Silvestrelli}, \citenamefont {Seif}, \citenamefont {Subashchandrabose},\
  and\ \citenamefont {Ambrosetti}}]{silvestrelli2022screening}%
  \BibitemOpen
  \bibfield  {author} {\bibinfo {author} {\bibfnamefont {Pier~Luigi}\
  \bibnamefont {Silvestrelli}}, \bibinfo {author} {\bibfnamefont
  {A}~\bibnamefont {Seif}}, \bibinfo {author} {\bibfnamefont {S}~\bibnamefont
  {Subashchandrabose}}, \ and\ \bibinfo {author} {\bibfnamefont {Alberto}\
  \bibnamefont {Ambrosetti}},\ }\bibfield  {title} {\enquote {\bibinfo {title}
  {Screening and antiscreening in fullerene-like cages: dipole-field
  amplification with ionic nanocages},}\ }\href@noop {} {\bibfield  {journal}
  {\bibinfo  {journal} {arXiv preprint arXiv:2210.05324}\ } (\bibinfo {year}
  {2022})}\BibitemShut {NoStop}%
\bibitem [{\citenamefont {Lu}\ \emph {et~al.}(2001)\citenamefont {Lu},
  \citenamefont {Zhou}, \citenamefont {Luo}, \citenamefont {Huang},
  \citenamefont {Zhang},\ and\ \citenamefont {Zhao}}]{lu2001structural}%
  \BibitemOpen
  \bibfield  {author} {\bibinfo {author} {\bibfnamefont {Jing}\ \bibnamefont
  {Lu}}, \bibinfo {author} {\bibfnamefont {Yunsong}\ \bibnamefont {Zhou}},
  \bibinfo {author} {\bibfnamefont {Yin}\ \bibnamefont {Luo}}, \bibinfo
  {author} {\bibfnamefont {Yuanhe}\ \bibnamefont {Huang}}, \bibinfo {author}
  {\bibfnamefont {Xinwei}\ \bibnamefont {Zhang}}, \ and\ \bibinfo {author}
  {\bibfnamefont {Xiangeng}\ \bibnamefont {Zhao}},\ }\bibfield  {title}
  {\enquote {\bibinfo {title} {Structural and electronic properties of
  heterofullerene c59p},}\ }\href@noop {} {\bibfield  {journal} {\bibinfo
  {journal} {Mol. Phys.}\ }\textbf {\bibinfo {volume} {99}},\ \bibinfo {pages}
  {1203--1207} (\bibinfo {year} {2001})}\BibitemShut {NoStop}%
\bibitem [{\citenamefont {Li}\ and\ \citenamefont
  {Jiao}(2009)}]{li2009theoretical}%
  \BibitemOpen
  \bibfield  {author} {\bibinfo {author} {\bibfnamefont {Xiao~Jun}\
  \bibnamefont {Li}}\ and\ \bibinfo {author} {\bibfnamefont {Geng~Sheng}\
  \bibnamefont {Jiao}},\ }\bibfield  {title} {\enquote {\bibinfo {title}
  {Theoretical studies of the functionalized derivatives of fullerene c24h24 by
  attaching a variety of chemical groups},}\ }\href@noop {} {\bibfield
  {journal} {\bibinfo  {journal} {Comput. Theor. Chem.}\ }\textbf {\bibinfo
  {volume} {893}},\ \bibinfo {pages} {26--30} (\bibinfo {year}
  {2009})}\BibitemShut {NoStop}%
\bibitem [{\citenamefont {Dheivamalar}\ and\ \citenamefont
  {Sugi}(2015)}]{dheivamalar2015density}%
  \BibitemOpen
  \bibfield  {author} {\bibinfo {author} {\bibfnamefont {S}~\bibnamefont
  {Dheivamalar}}\ and\ \bibinfo {author} {\bibfnamefont {L}~\bibnamefont
  {Sugi}},\ }\bibfield  {title} {\enquote {\bibinfo {title} {Density functional
  theory (dft) investigations on doped fullerene with heteroatom
  substitution},}\ }\href@noop {} {\bibfield  {journal} {\bibinfo  {journal}
  {Spectrochim. Acta A}\ }\textbf {\bibinfo {volume} {151}},\ \bibinfo {pages}
  {687--695} (\bibinfo {year} {2015})}\BibitemShut {NoStop}%
\bibitem [{\citenamefont {Lovas}\ \emph {et~al.}(2005)\citenamefont {Lovas},
  \citenamefont {Tiemann}, \citenamefont {Coursey}, \citenamefont
  {Kotochigova}, \citenamefont {Chang}, \citenamefont {Olsen},\ and\
  \citenamefont {Dragoset}}]{lovas2005nist}%
  \BibitemOpen
  \bibfield  {author} {\bibinfo {author} {\bibfnamefont {Frank~J.}\
  \bibnamefont {Lovas}}, \bibinfo {author} {\bibfnamefont {Eberhard}\
  \bibnamefont {Tiemann}}, \bibinfo {author} {\bibfnamefont {J.S.}\
  \bibnamefont {Coursey}}, \bibinfo {author} {\bibfnamefont {S.A.}\
  \bibnamefont {Kotochigova}}, \bibinfo {author} {\bibfnamefont
  {J.}~\bibnamefont {Chang}}, \bibinfo {author} {\bibfnamefont
  {K.}~\bibnamefont {Olsen}}, \ and\ \bibinfo {author} {\bibfnamefont {R.A.}\
  \bibnamefont {Dragoset}},\ }\bibfield  {title} {\enquote {\bibinfo {title}
  {Nist standard reference database 114},}\ }\href@noop {} {\bibfield
  {journal} {\bibinfo  {journal} {NIST Diatomic Spectral Database}\ } (\bibinfo
  {year} {2005})}\BibitemShut {NoStop}%
\bibitem [{\citenamefont {Shostak}\ \emph {et~al.}(1991)\citenamefont
  {Shostak}, \citenamefont {Ebenstein},\ and\ \citenamefont
  {Muenter}}]{shostak1991dipole}%
  \BibitemOpen
  \bibfield  {author} {\bibinfo {author} {\bibfnamefont {Shelley~L}\
  \bibnamefont {Shostak}}, \bibinfo {author} {\bibfnamefont {William~L}\
  \bibnamefont {Ebenstein}}, \ and\ \bibinfo {author} {\bibfnamefont {John~S}\
  \bibnamefont {Muenter}},\ }\bibfield  {title} {\enquote {\bibinfo {title}
  {The dipole moment of water. i. dipole moments and hyperfine properties of
  h2o and hdo in the ground and excited vibrational states},}\ }\href@noop {}
  {\bibfield  {journal} {\bibinfo  {journal} {J. Chem. Phys.}\ }\textbf
  {\bibinfo {volume} {94}},\ \bibinfo {pages} {5875--5882} (\bibinfo {year}
  {1991})}\BibitemShut {NoStop}%
\bibitem [{\citenamefont {Iouchtchenko}\ and\ \citenamefont
  {Roy}(2018)}]{iouchtchenko2018ground}%
  \BibitemOpen
  \bibfield  {author} {\bibinfo {author} {\bibfnamefont {Dmitri}\ \bibnamefont
  {Iouchtchenko}}\ and\ \bibinfo {author} {\bibfnamefont {Pierre-Nicholas}\
  \bibnamefont {Roy}},\ }\bibfield  {title} {\enquote {\bibinfo {title} {Ground
  states of linear rotor chains via the density matrix renormalization
  group},}\ }\href@noop {} {\bibfield  {journal} {\bibinfo  {journal} {J. Chem.
  Phys.}\ }\textbf {\bibinfo {volume} {148}},\ \bibinfo {pages} {134115}
  (\bibinfo {year} {2018})}\BibitemShut {NoStop}%
\bibitem [{\citenamefont {Latorre}\ and\ \citenamefont
  {Or{\'u}s}(2004)}]{latorre2004adiabatic}%
  \BibitemOpen
  \bibfield  {author} {\bibinfo {author} {\bibfnamefont {Jos{\'e}~Ignacio}\
  \bibnamefont {Latorre}}\ and\ \bibinfo {author} {\bibfnamefont {Rom{\'a}n}\
  \bibnamefont {Or{\'u}s}},\ }\bibfield  {title} {\enquote {\bibinfo {title}
  {Adiabatic quantum computation and quantum phase transitions},}\ }\href@noop
  {} {\bibfield  {journal} {\bibinfo  {journal} {Phys. Rev. A}\ }\textbf
  {\bibinfo {volume} {69}},\ \bibinfo {pages} {062302} (\bibinfo {year}
  {2004})}\BibitemShut {NoStop}%
\bibitem [{\citenamefont {Sch{\"u}tzhold}\ and\ \citenamefont
  {Schaller}(2006)}]{schutzhold2006adiabatic}%
  \BibitemOpen
  \bibfield  {author} {\bibinfo {author} {\bibfnamefont {Ralf}\ \bibnamefont
  {Sch{\"u}tzhold}}\ and\ \bibinfo {author} {\bibfnamefont {Gernot}\
  \bibnamefont {Schaller}},\ }\bibfield  {title} {\enquote {\bibinfo {title}
  {Adiabatic quantum algorithms as quantum phase transitions: First versus
  second order},}\ }\href@noop {} {\bibfield  {journal} {\bibinfo  {journal}
  {Phys. Rev. A}\ }\textbf {\bibinfo {volume} {74}},\ \bibinfo {pages}
  {060304(R)} (\bibinfo {year} {2006})}\BibitemShut {NoStop}%
\bibitem [{\citenamefont {Auciello}\ \emph {et~al.}(1998)\citenamefont
  {Auciello}, \citenamefont {Scott}, \citenamefont {Ramesh} \emph
  {et~al.}}]{auciello1998physics}%
  \BibitemOpen
  \bibfield  {author} {\bibinfo {author} {\bibfnamefont {Orlando}\ \bibnamefont
  {Auciello}}, \bibinfo {author} {\bibfnamefont {James~F}\ \bibnamefont
  {Scott}}, \bibinfo {author} {\bibfnamefont {Ramamoorthy}\ \bibnamefont
  {Ramesh}},  \emph {et~al.},\ }\bibfield  {title} {\enquote {\bibinfo {title}
  {The physics of ferroelectric memories},}\ }\href@noop {} {\bibfield
  {journal} {\bibinfo  {journal} {Phys. Today}\ }\textbf {\bibinfo {volume}
  {51}},\ \bibinfo {pages} {22--27} (\bibinfo {year} {1998})}\BibitemShut
  {NoStop}%
\bibitem [{\citenamefont {Damjanovic}\ \emph {et~al.}(2001)\citenamefont
  {Damjanovic}, \citenamefont {Muralt},\ and\ \citenamefont
  {Setter}}]{damjanovic2001ferroelectric}%
  \BibitemOpen
  \bibfield  {author} {\bibinfo {author} {\bibfnamefont {Dragan}\ \bibnamefont
  {Damjanovic}}, \bibinfo {author} {\bibfnamefont {Paul}\ \bibnamefont
  {Muralt}}, \ and\ \bibinfo {author} {\bibfnamefont {Nava}\ \bibnamefont
  {Setter}},\ }\bibfield  {title} {\enquote {\bibinfo {title} {Ferroelectric
  sensors},}\ }\href@noop {} {\bibfield  {journal} {\bibinfo  {journal} {IEEE
  Sens. J.}\ }\textbf {\bibinfo {volume} {1}},\ \bibinfo {pages} {191--206}
  (\bibinfo {year} {2001})}\BibitemShut {NoStop}%
\bibitem [{\citenamefont {Kim}\ \emph {et~al.}(2018)\citenamefont {Kim},
  \citenamefont {Kim},\ and\ \citenamefont {Kim}}]{kim2018application}%
  \BibitemOpen
  \bibfield  {author} {\bibinfo {author} {\bibfnamefont {Tae~Yun}\ \bibnamefont
  {Kim}}, \bibinfo {author} {\bibfnamefont {Sung~Kyun}\ \bibnamefont {Kim}}, \
  and\ \bibinfo {author} {\bibfnamefont {Sang-Woo}\ \bibnamefont {Kim}},\
  }\bibfield  {title} {\enquote {\bibinfo {title} {Application of ferroelectric
  materials for improving output power of energy harvesters},}\ }\href@noop {}
  {\bibfield  {journal} {\bibinfo  {journal} {Nano Convergence}\ }\textbf
  {\bibinfo {volume} {5}},\ \bibinfo {pages} {1--16} (\bibinfo {year}
  {2018})}\BibitemShut {NoStop}%
\bibitem [{\citenamefont {Krachmalnicoff}\ \emph {et~al.}(2016)\citenamefont
  {Krachmalnicoff}, \citenamefont {Bounds}, \citenamefont {Mamone},
  \citenamefont {Alom}, \citenamefont {Concistr{\`e}}, \citenamefont {Meier},
  \citenamefont {Kou{\v{r}}il}, \citenamefont {Light}, \citenamefont {Johnson},
  \citenamefont {Rols} \emph {et~al.}}]{krachmalnicoff2016dipolar}%
  \BibitemOpen
  \bibfield  {author} {\bibinfo {author} {\bibfnamefont {Andrea}\ \bibnamefont
  {Krachmalnicoff}}, \bibinfo {author} {\bibfnamefont {Richard}\ \bibnamefont
  {Bounds}}, \bibinfo {author} {\bibfnamefont {Salvatore}\ \bibnamefont
  {Mamone}}, \bibinfo {author} {\bibfnamefont {Shamim}\ \bibnamefont {Alom}},
  \bibinfo {author} {\bibfnamefont {Maria}\ \bibnamefont {Concistr{\`e}}},
  \bibinfo {author} {\bibfnamefont {Benno}\ \bibnamefont {Meier}}, \bibinfo
  {author} {\bibfnamefont {Karel}\ \bibnamefont {Kou{\v{r}}il}}, \bibinfo
  {author} {\bibfnamefont {Mark~E}\ \bibnamefont {Light}}, \bibinfo {author}
  {\bibfnamefont {Mark~R}\ \bibnamefont {Johnson}}, \bibinfo {author}
  {\bibfnamefont {St{\'e}phane}\ \bibnamefont {Rols}},  \emph {et~al.},\
  }\bibfield  {title} {\enquote {\bibinfo {title} {The dipolar endofullerene
  hf@ c60},}\ }\href@noop {} {\bibfield  {journal} {\bibinfo  {journal} {Nat.
  Chem.}\ }\textbf {\bibinfo {volume} {8}},\ \bibinfo {pages} {953--957}
  (\bibinfo {year} {2016})}\BibitemShut {NoStop}%
\bibitem [{\citenamefont {Wespiser}\ \emph {et~al.}(2022)\citenamefont
  {Wespiser}, \citenamefont {Putaud}, \citenamefont {Kalugina}, \citenamefont
  {Soldera}, \citenamefont {Roy}, \citenamefont {Michaut},\ and\ \citenamefont
  {Ayotte}}]{wespiser2022ro}%
  \BibitemOpen
  \bibfield  {author} {\bibinfo {author} {\bibfnamefont {Cl{\'e}ment}\
  \bibnamefont {Wespiser}}, \bibinfo {author} {\bibfnamefont {Thomas}\
  \bibnamefont {Putaud}}, \bibinfo {author} {\bibfnamefont {Yulia}\
  \bibnamefont {Kalugina}}, \bibinfo {author} {\bibfnamefont {Armand}\
  \bibnamefont {Soldera}}, \bibinfo {author} {\bibfnamefont {Pierre-Nicholas}\
  \bibnamefont {Roy}}, \bibinfo {author} {\bibfnamefont {Xavier}\ \bibnamefont
  {Michaut}}, \ and\ \bibinfo {author} {\bibfnamefont {Patrick}\ \bibnamefont
  {Ayotte}},\ }\bibfield  {title} {\enquote {\bibinfo {title} {Ro-translational
  dynamics of confined water. i. the confined asymmetric rotor model},}\
  }\href@noop {} {\bibfield  {journal} {\bibinfo  {journal} {J. Chem. Phys.}\
  }\textbf {\bibinfo {volume} {156}},\ \bibinfo {pages} {074304} (\bibinfo
  {year} {2022})}\BibitemShut {NoStop}%
\bibitem [{\citenamefont {Dolgonos}\ and\ \citenamefont
  {Peslherbe}(2014)}]{dolgonos2014encapsulation}%
  \BibitemOpen
  \bibfield  {author} {\bibinfo {author} {\bibfnamefont {Grygoriy~A}\
  \bibnamefont {Dolgonos}}\ and\ \bibinfo {author} {\bibfnamefont {Gilles~H}\
  \bibnamefont {Peslherbe}},\ }\bibfield  {title} {\enquote {\bibinfo {title}
  {Encapsulation of diatomic molecules in fullerene c 60: implications for
  their main properties},}\ }\href@noop {} {\bibfield  {journal} {\bibinfo
  {journal} {Phys. Chem. Chem. Phys.}\ }\textbf {\bibinfo {volume} {16}},\
  \bibinfo {pages} {26294--26305} (\bibinfo {year} {2014})}\BibitemShut
  {NoStop}%
\bibitem [{\citenamefont {Coleman}\ and\ \citenamefont
  {Schofield}(2005)}]{coleman2005quantum}%
  \BibitemOpen
  \bibfield  {author} {\bibinfo {author} {\bibfnamefont {Piers}\ \bibnamefont
  {Coleman}}\ and\ \bibinfo {author} {\bibfnamefont {Andrew~J}\ \bibnamefont
  {Schofield}},\ }\bibfield  {title} {\enquote {\bibinfo {title} {Quantum
  criticality},}\ }\href@noop {} {\bibfield  {journal} {\bibinfo  {journal}
  {Nature}\ }\textbf {\bibinfo {volume} {433}},\ \bibinfo {pages} {226--229}
  (\bibinfo {year} {2005})}\BibitemShut {NoStop}%
\bibitem [{\citenamefont {Sachdev}\ and\ \citenamefont
  {Keimer}(2011)}]{sachdev2011quantumcrit}%
  \BibitemOpen
  \bibfield  {author} {\bibinfo {author} {\bibfnamefont {Subir}\ \bibnamefont
  {Sachdev}}\ and\ \bibinfo {author} {\bibfnamefont {Bernhard}\ \bibnamefont
  {Keimer}},\ }\bibfield  {title} {\enquote {\bibinfo {title} {Quantum
  criticality},}\ }\href@noop {} {\bibfield  {journal} {\bibinfo  {journal}
  {Phys. Today}\ }\textbf {\bibinfo {volume} {64}},\ \bibinfo {pages} {29}
  (\bibinfo {year} {2011})}\BibitemShut {NoStop}%
\bibitem [{\citenamefont {Hauser}\ and\ \citenamefont
  {Pototschnig}(2022)}]{hauser2022vibronic}%
  \BibitemOpen
  \bibfield  {author} {\bibinfo {author} {\bibfnamefont {Andreas~W}\
  \bibnamefont {Hauser}}\ and\ \bibinfo {author} {\bibfnamefont {Johann~V}\
  \bibnamefont {Pototschnig}},\ }\bibfield  {title} {\enquote {\bibinfo {title}
  {Vibronic coupling in spherically encapsulated, diatomic molecules:
  Prediction of a renner--teller-like effect for endofullerenes},}\ }\href@noop
  {} {\bibfield  {journal} {\bibinfo  {journal} {J. Phys. Chem. A}\ }\textbf
  {\bibinfo {volume} {126}},\ \bibinfo {pages} {1674--1680} (\bibinfo {year}
  {2022})}\BibitemShut {NoStop}%
\end{thebibliography}
%

\end{document}